\documentclass[twocolumn,prl,showpacs,preprintnumbers,amsmath]{revtex4}
\usepackage{graphicx}
\usepackage{dcolumn}
\usepackage{bm}

\begin{document}

\title{Spectroscopic Signatures of Massless Gap Opening in Graphene}
\author{L. Benfatto$^{1,2,3}$}

\email{lara.benfatto@roma1.infn.it}
\author{E. Cappelluti$^{2,3}$}

\email{emmanuele.cappelluti@roma1.infn.it}
\affiliation{$^1$Centro Studi e Ricerche ``Enrico Fermi'',
via Panisperna 89/A, I-00184,
Rome, Italy}

\affiliation{$^2$SMC-INFM, CNR-INFM, via dei
Taurini 19, 00185 Roma, Italy}

\affiliation{$^3$Dipart. di Fisica, Universit\`a ``La Sapienza'',
P.le A. Moro 2, 00185 Roma, Italy}

\begin{abstract}
Gap opening in graphene is usually discussed in terms of a
semiconducting-like spectrum, where the appearance of a finite gap at the
Dirac point is accompanied by a finite mass for the fermions. In this
letter we propose a gap scenario from graphene which preserves the massless
characters of the carriers. This approach explains recent spectroscopic
measurements carried out in epitaxially-grown graphene, ranging from 
photoemission to optical transmission.
\end{abstract}

\maketitle

In the last few years the realization of free-standing single layers of
graphene opened a new route to the possibility of investigating the
behavior of massless Dirac fermion in two dimensions. Indeed, in graphene
the valence and conduction bands are formed by the $p_z$ orbitals of the
carbon atoms arranged on the two sublattices A and B of the honeycomb
lattice. When the two sublattices are electrostatically equivalent the two
bands meet at the K (K$'$) points of the Brillouin zone, leading to a
zero-gap semiconductor with two conical bands $\varepsilon_k^\pm=\pm v_{\rm
F}k$ ($k=|{\bf k}|$), which resemble relativistic Dirac carriers with zero
mass and Fermi velocity $v_F$ . For this reason, K and K$'$ are usually
referred to as Dirac points.  Even though the massless Dirac spectrum makes
graphene the perfect playground for investigating relativistic effects in
quantum systems, for device application a tunable-gap semiconducting
behavior would be more suitable.  Along this perspective, of remarkable
interest are some recent experiments of Angle Resolved PhotoEmission
Spectroscopy (ARPES) in epitaxially-grown graphene, which reveal a finite
band splitting $2\Delta \sim 0.26$ eV at the K point
\cite{lanzara_natmat07,zhou_condmat08}. In these works the authors propose
that the splitting arises from the inequivalence of the A and B
sublattices, which in turn leads to a massive (ms) gapped spectrum
\cite{gusynin_review2007,vozmediano_prb07}:
\begin{equation}
E_{k,\pm}^{\rm ms}=\pm \sqrt{\left(v_{\rm F}k\right)^2+\Delta^2}.
\label{gap_mass}
\end{equation}
The magnitude of the band-splitting gap has been reported to decrease by
increasing the number of layers, or, from another perspective, by reducing
the induced charge density. Similar ARPES spectra were reported previously
in Ref.\ \cite{rotenberg_natphys07}, although a different interpretation was
proposed\cite{rotenberg_natphys07,rotenberg_njp07,comments}. The most
convincing argument in such a controversy comes from the electronic
dispersion far from the Dirac point, which is at odd with the massive
gapped spectrum of Eq.\ (\ref{gap_mass}).  Indeed, Eq.\ (\ref{gap_mass})
predicts $E_{k,\pm}^{\rm ms}$ to be unaffected by the gap opening in the
$|E_{k,\pm}^{\rm ms}| \gg \Delta$ regime, and in particular the linear
asymptotic behavior of the upper band $E_{k,+}^{\rm ms} \approx v_{\rm F}k$
($E_{k,+}^{\rm ms}\gg \Delta$) should match the linear behavior of the
lower band $E_{k,-}^{\rm ms} \approx -v_{\rm F}k$.  A careful analysis of
the ARPES data reveals on the contrary a finite off-shift of the two
asymptotic linear behaviors
\cite{lanzara_natmat07,zhou_condmat08,rotenberg_natphys07,rotenberg_njp07},
which cannot be explained even by the periodic modulation induced by the
substrate \cite{kim_cm08}.  Further discrepancies appear also in the profile
of the dispersion at finite $k_x$ away from the K point
(Ref.\ \cite{lanzara_natmat07}, Fig.\ S3): instead of the parabolic shape
predicted by Eq.\ (\ref{gap_mass}), a ``V'' shaped conic profile appears,
characteristic of a massless dispersion.

Quite remarkably, the issue of the gap opening has been raised also by very
recent optical-absorption measurements in epitaxial
graphene\cite{dawlaty}. Indeed, the analysis of the optical spectra based
on the massive gap model (\ref{gap_mass}) has two major drawbacks: from one
side, it would suggest that no gap (or a negligible one) is present in the
system, in contrast with ARPES results, from the other side it fails in
reproducing the data in the visible frequency range, where the
``universal'' conductivity value of $e^2/4\hbar$ is expected
theoretically\cite{ando_jpsj02,sharapov_prl06,peres_cm08}, and tested
experimentally in multi-layer graphite samples\cite{kuzmenko_prl08} or
few-layers suspended graphene\cite{geim08}. In this Letter, we propose a
different gap scenario that reconciles the gapped nature of the spectrum
with the massless character of the fermions in graphene.  As we shall see,
such a gap model not only accounts very well for the ARPES spectra, but
reconciles also optical measurements performed from the visible to the
far-infrared (IR) regime.

Let us start by introducing the Hamiltonian for free electrons in the
graphene honeycomb lattice, in terms of the usual spinor $\psi_{\bf
k}^\dagger=(c_{{\bf k},{\rm A}}^\dagger, c_{{\bf k},{\rm B}}^\dagger)$.
Linearizing around the K Dirac point (we put $\hbar=1$), and using
${\bf k}=k(\cos\phi,\sin\phi)$, we can write:
\begin{eqnarray}
\hat{H}^0_{\bf k}
=
v_{\rm F}k\left(
\begin{array}{cc}
0 & \mbox{e}^{-i\phi} \\
\mbox{e}^{i\phi} & 0
\end{array}
\right),
\end{eqnarray}
whose eigenvalues are the usual gapless Dirac cones $\varepsilon_k^\pm=\pm
v_{\rm F}k$.  If the A and B sublattices are electrostatically
inequivalent, an additional term $\propto \Delta \hat{\sigma}_z$
($\hat\sigma_{i=I,x,y,z}$ being the Pauli matrices) should be added to
$\hat H^0_{\bf k}$, and the gapped spectrum of Eq. (\ref{gap_mass}) is
recovered. In similar way, in Ref.\ \cite{vozmediano_prb07} it was shown
that a gap opening, related to inter-valley scattering, is always associated
with a mass onset.  
Finally, off-diagonal intra-valley processes of the kind
$Q_1^x\hat{\sigma}_x+Q_1^y\hat{\sigma}_y$ were claimed to lead to a mere
displacement of the Dirac cone off the K (K$'$) points
\cite{vozmediano_prb07}.  However, this statement is correct only as far as
a constant value of $Q_1^x$, $Q_1^y$ for ${\bf k}\rightarrow 0$ is assumed,
while in general a momentum dependence of the off-diagonal self-energy
cannot be ruled out.  For instance, the (unscreened) Coulomb interaction
leads to the off-diagonal self-energy $\hat{\Sigma}_{\bf k} \propto
k\log(k_c/k)[\cos\phi\hat{\sigma}_x+\sin\phi\hat{\sigma}_y]$, where $k_c$
is a momentum cut-off for the Dirac-like conical
behavior\cite{mishchenko_prl07}.  In this case, a correction to the linear
Dirac dispersion is achieved, but no gap opens because $\hat{\Sigma}$
vanishes as $k\rightarrow 0$.

In this
Letter we want to explore a somehow intermediate possibility, where the
off-diagonal self-energy has the structure:
\begin{equation}
\hat{\Sigma}_{\bf k}
\simeq
\Delta [\cos\phi\hat{\sigma}_x+\sin\phi\hat{\sigma}_y].
\label{selfoff}
\end{equation}
It is straightforward to check that, in this case, a gap is opened at the K
(K$'$) points without the onset of a massive term. Indeed,
Eq. (\ref{selfoff}) leads to the massless (ml) gapped spectrum of the form:
\begin{eqnarray}
E^{\rm ml}_{k,\pm}=\pm(v_{\rm F}k+\Delta),
\label{gap_nomass}
\end{eqnarray}
with a corresponding density of states (DOS):
\begin{eqnarray}
N(\omega) =\frac{V_{\rm BZ}}{2\pi v_{\rm F}^2}(|\omega|-\Delta)
\theta(|\omega|-\Delta),
\label{dos_nomass}
\end{eqnarray}
where $V_{\rm BZ}=5.24$ \AA$^2$ is the volume of the Brillouin zone.

\begin{figure}[t]
\includegraphics[scale=0.34]{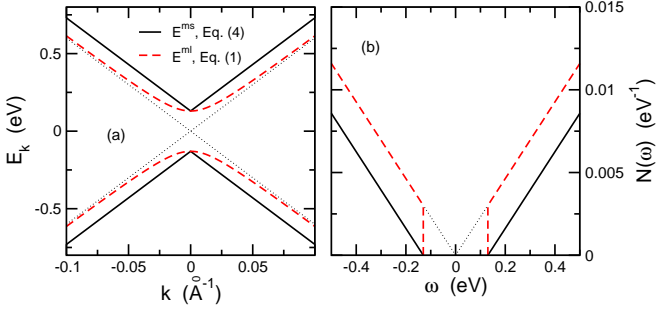}
\caption{(Color online). Electronic dispersion (a) and DOS (b) for the massless
$E^{\rm ml}$ (solid black line) and massive $E^{\rm ms}$
(dashed red line) gapped
models, with $\Delta=0.13$ eV.
The dotted line refers to the Dirac-like dispersion $\varepsilon$ 
in the absence of any gap.}
\label{f-ek-dos}
\end{figure}

While the main aim of this Letter is to focus on the phenomenological
outcomes of this spectrum, a brief discussion on the microscopic origin of
the self-energy (\ref{selfoff}) will be given later.  In Fig.
\ref{f-ek-dos} we compare the electronic dispersion (panel a) and the DOS
(panel b) of the models (\ref{gap_nomass}) and
(\ref{gap_mass}).  Two main striking features need here to be stressed:
$i$) the gap in Eq. (\ref{gap_nomass}), as mentioned, is created without
affecting the conical electronic dispersion.  In this sense no massive term
is induced; $ii$) in contrast to the case of the massive gap model
(\ref{gap_mass}), the lower and upper cones of the spectrum
(\ref{gap_mass}) are misaligned by the quantity $2\Delta$. Such anomalous
features are reflected  in the DOS [Eq. (\ref{dos_nomass})] where the
linear vanishing of $N(\omega)$ as $|\omega|\rightarrow \Delta^+$ in the
model (\ref{gap_nomass}) points out the absence of a quadratic massive
term, while the misalignment of the cones is reflected in a corresponding
mismatch of the linear dependence of the DOS, which does not extrapolate to
zero at the Dirac point.

\begin{figure}[t]
\includegraphics[angle=0,scale=0.3]{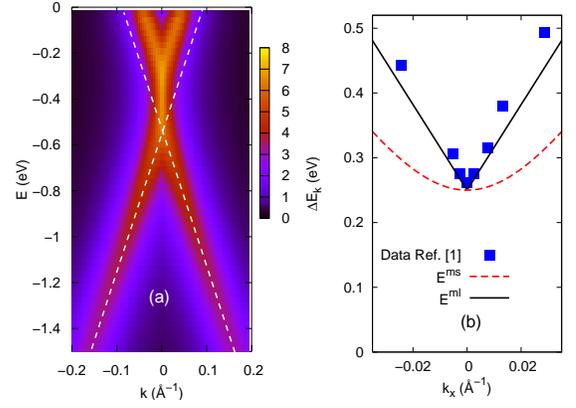}
\caption{(color online). (a) Intensity map of the spectral function
$A(k,\omega)$ of the massless model [Eq. (\ref{gap_nomass})]
for $\Delta=0.13$ eV,
$v_F=6$ eV \AA and $\Gamma$ taken from the experiments (see text). Dashed
white lines show the mismatch between the asymptotic linear behavior of the
upper and lower bands.  (b) Plot of the gap edge at finite $k_x$ for the
massless and massive gap model ($\Delta, v_F$ as in panel (a)). Dark
squares are experimental data taken from Ref. \cite{lanzara_natmat07}.}
\label{f-arpes}
\end{figure}

To make a comparison with the ARPES results in epitaxially-grown graphene,
we show in Fig. \ref{f-arpes}a the calculated spectral intensity
$I({\bf k},\omega)=A({\bf k},\omega)f(\omega)$
of the massless model $E_\pm^{\rm ml}$,
where $\mu$ is the chemical potential, $f(\omega)$ is the Fermi function
and $A({\bf k},\omega)$ is the spectral function
$A({\bf k},\omega)=-\mbox{Tr}\{\mbox{Im}[(\omega+\mu+i\Gamma)\hat{I}
-\hat{H}_{\bf k}^0-\hat{\Sigma}_{\bf k}]^{-1}\}=
\Gamma/(2\pi)\{[(\omega-\mu-v_F k-\Delta)^2+\Gamma^2]^{-1}+ 
[(\omega-\mu+v_F k+\Delta)^2+\Gamma^2]^{-1}\}$.
In Fig.\ \ref{f-arpes} we used $\mu=0.4$ eV, $2\Delta =0.26$ eV ($\Delta =
0.13$ eV)\cite{lanzara_natmat07} and we assumed a quasi-particle scattering
rate $\Gamma(\omega) = \Gamma_0 +\alpha |\omega|$, with $\Gamma_0=0.165$ eV
and $\alpha=0.11$ fitted from the ARPES data\cite{zhou_condmat08} away from
the Dirac point using $v_{\rm F}=6$ eV\, \AA \cite{guinea_review}.
Similar values could be estimated from Ref. \cite{rotenberg_natphys07}.
Notice that the presence of such a large scattering time at the Dirac point
partly spoils the gap feature as $k\rightarrow 0$, because the two peaks of
$A({\bf k}=0,\omega)$ at $\omega=\pm \Delta$ overlap and cannot be
resolved, leading to the controversial interpretation of similar ARPES
spectra in Refs.\ \cite{rotenberg_natphys07,rotenberg_njp07} and
\cite{lanzara_natmat07,zhou_condmat08}, respectively. The same feature is
observed in Fig.\ \ref{f-arpes}a, where the exact dispersion at the Dirac
point is not detectable due to the large electronic damping. Nevertheless,
one can still clearly resolve the net misalignment between the upper and
lower Dirac cones, which is peculiar to the model (\ref{gap_nomass}), and
it is in perfect agreement with all the existing experimental data
\cite{lanzara_natmat07,rotenberg_natphys07,zhou_condmat08}. Moreover, ARPES
data taken away from the K point provide us also with a direct evidence of
the linear (massless) behavior of the electronic dispersion at finite
$k$. By measuring the dispersion at $k_y=0$ and finite $k_x$, as extracted
from several $k_y$ cuts through the K point, one can easily discriminate
between the two models (\ref{gap_mass}) and (\ref{gap_nomass}). Indeed,
within the model (\ref{gap_mass}) the energy spectrum
$\sqrt{\Delta^2+(v_F k_x)^2}$ would appear parabolic within a momentum window
$k_x\lesssim \Delta/v_F \simeq 0.02$ \,\AA$^{-1}$, while within the massless
model (\ref{gap_nomass}) one expects a linear increase of the gap,
$\Delta+v_F|k_x|$. The experimental data by Ref. \cite{lanzara_natmat07}
(Fig.\ S3 in Supplementary material) are reported in Fig. \ref{f-arpes}b:
as one can see, the massive model $E_{\pm}^{\rm ms}$ shows no resemblance
with the data, while the massless model $E^{\rm ml}_\pm$ allows one an
excellent fit of the dispersion without any adjustable parameter.

\begin{figure}[t]
\includegraphics[angle=0,scale=0.4]{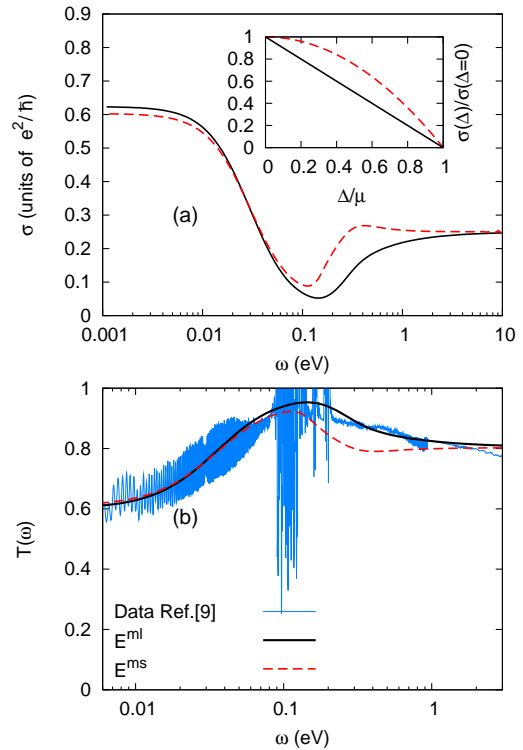}
\caption{(Color online). (a) Optical conductivity $\sigma(\Omega)$ for
$\mu=0.1$ eV, $T=300$ K, $\Gamma=15$ meV for the massless gap model
(solid black line, $\Delta=45$ meV) and for the massive gap model (dashed
red line, $\Delta=73$ meV). Inset: reduction of the zero-frequency optical
conductivity in the two cases, as a function of $\Delta/\mu$. (b):
Comparison of the two models with the experimental data for $T(\omega)$
from Ref.\ \cite{dawlaty}.}
\label{f-opt}
\end{figure}

The massless character of the spectrum has also significant consequences on
the structure of the optical conductivity $\sigma(\Omega)$. To elucidate this
issue we evaluate here the optical conductivity in the bare-bubble
approximation. As usual, $\sigma(\Omega)$ is given by two parts, associated
to intraband and interband transitions\cite{ando_jpsj02,sharapov_prl06}.
At the leading order in $\Gamma/|\mu|, \Gamma/\Delta \ll 1$, we obtain:
\begin{eqnarray}
\sigma_{\rm intra}(\Omega)&=&
-\frac{e^2}{\pi\hbar}
\frac{2\Gamma}{\Omega^2+4\Gamma^2}
\nonumber\\
&&\times
\int d\omega
\frac{df(\omega-\mu)}{d\omega}
\left(|\omega|-\Delta\right)
\theta[|\omega|-\Delta]
\nonumber\\
&\approx\atop{T\rightarrow 0}&
\frac{e^2}{\hbar}\delta(\Omega)
\left(|\mu|-\Delta\right)
\theta[|\mu|-\Delta],
\label{intra}
\\
\sigma_{\rm inter}(\Omega)
&=&
\frac{e^2}{\pi\hbar}
\int d\omega
\frac{f(\omega-\mu)-f(\omega+\Omega-\mu)}{4\Omega}
\nonumber\\
&&\times
\frac{2\Gamma}{(\omega+\Omega/2)^2+4\Gamma^2}
\left(|\Omega|-2\Delta\right)
\theta\left[|\Omega|-2\Delta\right]
\nonumber\\
&\approx\atop{T\rightarrow 0}&
\frac{e^2}{4\hbar}
\left(1-\frac{2\Delta}{|\Omega|}\right)
\theta\left[|\Omega|-2\mbox{max}\left(\Delta,|\mu|\right)\right].
\label{inter}
\end{eqnarray}
As one can see, the general structure of the optical conductivity is the
same for all the gapped and ungapped models: a Drude peak of width
$\Gamma_{\rm opt}=2\Gamma$ and an interband contribution which starts at a
threshold given by the larger between $2\mu$ and $2\Delta$, and saturates at
$\Omega\gg\Delta,|\mu|$ to the universal value $e^2/4\hbar$
\cite{ando_jpsj02,sharapov_prl06,peres_cm08}. When a gap opens, part of the
spectral weight is transferred to the interband transitions and the Drude
peak decreases with respect to the ungapped case.
In the model (\ref{gap_mass}), $\sigma(\Omega)$ can be obtained
by replacing the factors $(|\mu|-\Delta)\rightarrow (|\mu|-\Delta^2/|\mu|)$
and $(1-2\Delta/|\Omega|) \rightarrow (1+4\Delta^2/|\Omega|^2)$
respectively in Eqs. (\ref{intra}) and
(\ref{inter})\cite{ando_jpsj02,sharapov_prl06}.
Two features
allow thus one to differentiate the models (\ref{gap_mass}) and
(\ref{gap_nomass}): the relative weight of interband and intraband
contributions and the shape of the conductivity at the interband
threshold. In particular, one can
see that ($i$) the gap-induced reduction of the Drude peak is much stronger
in the massless gap model (of order $\sim \Delta$) than in the massive one
(of order $\sim \Delta^2/|\mu|$), see inset of Fig. \ref{f-opt}a); ($ii$)
while the massive gap model would give rise to a {\em peak} above the
asymptotic value $e^2/4\hbar$ at the edge of the interband spectrum
and to a rapid saturation to the asymptotic value,
the massless model predicts a {\em depletion} in the correspondence of such
edge (due to the factor $1-2\Delta/\Omega<1$), followed by a slow
saturation to the universal value, see Fig. \ref{f-opt}a. Such a depletion
is again a consequence of the vanishing of the DOS at the gap in the
massless model (see Fig.\ \ref{f-ek-dos}b), with consequent reduction of
the spectral weight for transitions occurring between the two bands.

The present results open thus a new perspective in the analysis of the
optical properties of graphene grown epitaxially on SiC substrates, that
can be deduced from measurements of the optical transmission $T(\omega)$,
which in first approximation is related to the real part of the
single-layer optical conductivity as $T(\omega)=[1+N\sigma(\omega)
\sqrt{\mu_0/\epsilon_0}/(1+n_{\rm SiC})]^{-2}$\cite{dawlaty,peres_cm08},
where $n_{\rm SiC}$ is the refractive index of SiC and $N$ is the number of
layers. In Ref.\ \cite{dawlaty}, indeed, transmission data for few-layers
graphene samples in the frequency range from the far-IR to the visible were
analyzed within the context of the massive gap model described in Eq.\
(\ref{gap_mass}), and the gap was concluded to be negligible within the
experimental accuracy $\Delta \lesssim \Gamma_{\rm opt} \simeq 10$ meV
\cite{dawlaty}. Such a fit reproduces the data in the far-IR to mid-IR
range, but fails in the visible range, where, according to this fit, the
data would then indicated a conductivity {\em larger} than the universal
value $e^2/4\hbar$. The failure of the fit follow from the fact that the
size of optical transmission $T(\Omega)$ at $\Omega \approx 3$ eV and for
$\Omega \rightarrow 0$, and hence the size of the optical conductivity in
the corresponding range, are found to be of the same magnitude. According
to the previous discussion, this could be achieved by a transfer of
spectral weight from the Drude peak to the interband conductivity, as due
to a gap opening. However, to reproduce this feature within the massive gap
model one would need $\Delta \approx |\mu|\sqrt{1-\pi \Gamma/2|\mu|} = 87$
meV, where $|\mu| \simeq 0.1$ eV, $\Gamma=\Gamma_{\rm opt}/2 \approx 15$
meV are extracted from the interband edge and from the width of the Drude
peak \cite{dawlaty}.  With such large value, $\Delta \gg \Gamma$, the
massive gap would be clearly detectable as a sharp peak at the interband
edge, which is instead absent in the data. For this reason, the authors of
Ref.\ \cite{dawlaty} extracted a vanishing gap from the fit based on the
model (\ref{gap_mass}), which then fails in reproducing the data in the
visible range.

Such ambiguity can be naturally solved within the context of the massless
gap model where: $i$) a relatively smaller gap value is needed to make the
low and high frequency optical conductivity of the same magnitude; $ii$)
the opening of a massless gap does not give rise to any peak at the
threshold of interband transitions but to a {\em depletion} of the
conductivity with respect to the universal value, and then to a slow and
smooth crossover toward the high-frequency regime, as observed in the data.
This picture is in very good agreement with the actual experimental
measurements.  In Fig. \ref{f-opt}b we show the best fit to one set of the
experimental data of Ref. \cite{dawlaty} by using the massless gap model.
We take $|\mu| = 0.1$ eV, $T=300$ K and $\Gamma=15$ meV from the
experiments themselves and we estimate $\Delta=45$ meV, $N=18$. For
comparison, the fit with the massive gap model, constrained to reproduce
the experimental values of $T(\omega)$ in the low and high frequency limit,
would give $\Delta=73$ meV and $N=18$, and would result in a clear peak
(shoulder) at the interband edge.  We would like to stress that the even
though the absence on an interband peak can also be accounted for by a
vanishing gap, the similar magnitude of the low and high-frequency values
of the optical transmission is a clear indication of a reduced Drude
height, and hence of the presence of a gap.

So far, we have analyzed in details the consequences on ARPES and optical
spectra of the massless gap induced by a self-energy as described in Eq.\
(\ref{selfoff}).  In the last part of this Letter we discuss the possible
physical origin of such off-diagonal self-energy. To this aim we expand the
electron Green's function and the self-energy in terms of their Pauli
components where, for the case of equivalent carbon sublattice, we can
neglect the $\hat{\sigma}_z$ term.  In the noninteracting case the
off-diagonal Green's function, for instance $\propto \hat{\sigma}_x$, has
the simple form: $G_x({\bf
k},i\omega_n)=-k\cos\phi/[(i\omega_n+\mu)^2-v_{\rm F}^2k^2]$.  At the
Hartree-Fock level, one can quite generally write $\hat \Sigma_{\bf k}=
T\sum_{{\bf k}',n}V({\bf k}-{\bf k}')\hat G({\bf k},
i\omega_{n})\mbox{e}^{i\omega_n0^+}$, which can have off-diagonal
components depending on the form of the potential $V({\bf k}-{\bf k}')$
(see for example the case of unscreened Coulomb interaction discussed in
Ref.\ \cite{mishchenko_prl07}). While a completely momentum-independent
interaction would thus lead to a vanishing off-diagonal contribution, due
to the angular average, a small anisotropy in the scattering angle can lead
to a self energy of the form (\ref{selfoff}). For example, one can use
$V({\bf k}-{\bf k}')=V\theta(|\phi-\phi'|-\phi_c)$, where $\phi_c=\pi$
corresponding to isotropic scattering. In this case, one can easily check
that $\hat\Sigma_{\bf k}$ has exactly the form (\ref{selfoff}), with
$\Delta= V^2 k_c^2 \sin^2\phi_c/8\pi^2$.  Further investigation on the
possible source of anisotropic scattering could thus shed more light on the
microscopic origin of the proposed self-energy (\ref{selfoff}). In this
context, it is worth noting that a possible role of the doping is suggested
by the observed vanishing of both the gap and the band misalignment as a
function of the charge
concentration\cite{lanzara_natmat07,rotenberg_natphys07}.

In summary, we propose a gapped model for graphene which allows one to
reconcile the massless Dirac character of the carriers with the effects
related to a gap opening at the Dirac point.  We show that both ARPES and
optical-conductivity measurements give clear indications of such massless
gap opening in epitaxially-grown graphene.  Since ARPES measurements are
available only for epitaxially-grown graphene, we cannot rule out the
possibility that such a gap opening is restricted to these systems. Recent
tunneling data on epitaxially-grown\cite{vitali_cm08} and suspended
graphene \cite{andrei_cm08} are not conclusive on this respect: however,
our predictions can be further tested experimentally and, if confirmed,
they would pose stringent constraints on the interaction mechanisms at play
in graphene.

\begin{acknowledgments}
We acknowlegde useful discussions with L.~Boeri, A.~Kuzmenko,
S.~G.~Sharapov. We thanks the authors of Ref.\ \cite{dawlaty} for providing
us with the experimental data.
\end{acknowledgments}

\end{document}